\documentclass{amsart}
\usepackage{amsmath,float,amssymb,amsfonts}
\usepackage{url,cite}
\usepackage{multirow}
\usepackage{longtable}
\usepackage{xcolor}
\usepackage{verbatim}
\usepackage[linesnumbered,ruled,lined]{algorithm2e}
\usepackage{epsfig}
\usepackage{graphicx}
\usepackage{color}
\usepackage[english]{babel}
\usepackage{enumerate}

\newcommand{\pf}{{\bf Proof: }}
\newtheorem{definition}{Definition}
\newtheorem{theorem}{Theorem}
\newtheorem{lemma}{Lemma}

\newtheorem{corollary}{Corollary}
\newtheorem{case}{Case}
\newtheorem{rem}{Remark}
\usepackage{graphicx,tikz,pgf}
\usepackage{enumerate}
\usepackage{tikz}
\usepackage{verbatim}
\usetikzlibrary{calc,trees,positioning,arrows,chains,shapes.geometric,%
    decorations.pathreplacing,decorations.pathmorphing,shapes,%
    matrix,shapes.symbols}

\date{}

\author{Madhuparna Das$^{(1)}$}
\address[1]{}
\email{amimadhuparna@gmail.com}
\author{Goutam Paul$^{(2)}$}
\address[2]{Cryptology and Security Research Unit\\
        R. C. Bose Centre for Cryptology and Security\\
        Indian Statistical Institute, Kolkata 700108, India.}
\email{goutam.paul@isical.ac.in}

\begin{document}

\title{Analysis of the Game ``2048" and It's Generalization in Higher Dimensions}
\maketitle

\begin{abstract}
We theoretically analyze the popular mobile app game `2048' for the first time in $n$-dimensional space. We show that one can reach the maximum value $2^{n_1n_2+1}$ and $2^{\left({\prod_{i=1}^{d} n_i}\right)+1}$ for the two dimensional $n_1\times n_2$ board and $d$ dimensional $n_1\times n_2\times \ldots \times n_d$ board respectively.  We also present a strategy for the computer and a winning strategy for the human player in certain conditions.
\end{abstract}

\medskip


\section{Introduction}
`2048'~\cite{gameref} is a famous tile-mashing game created by Italian web developer Gabriele Cirulli and is extremely popular as a game app in mobile phones. In 2015's computer olympiad, competitors were given to play this game. The game's objective is to slide numbered tiles on a grid to combine them to create a tile with the number 2048. However, one can keep playing the game, creating tiles with larger numbers (such as a 32,768 tile)~\cite{cite1}. The game board begins with an initial configuration of two tiles of value 2 or 4, placed at arbitrary locations on the grid. The player then selects to move {\bf UP, DOWN, LEFT or RIGHT}, denoted as $\Uparrow, \Downarrow, \Leftarrow, \Rightarrow$ respectively.

\medskip

In this paper, we give a winning strategy and analyze the maximum value the player can reach for the standard two-dimensional board. Here a question arises that How can a player move the tiles if there is $d$ dimensional array that is there are $d$ many possibilities for each tile instead of 4 possibilities (which exists for the usual two-dimensional board)? The answer to this question is nothing but the generalization of the game 2048. We have developed a playing strategy for the player for the generalized version of 2048. 

\medskip

When the player has no legal moves (there are no empty spaces and no adjacent tiles with the same value), the game ends. Cirulli himself described 2048 as a clone of Veewo Studios' app 1024, who has said in the description of the app to be a clone of Threes!. 2048 became a viral hit~\cite{cite2, cite3}. The game has been described by the Wall Street Journal as ``almost like Candy Crush for math geeks"~\cite{cite4}, and Business Insider called it ``Threes on steroids"~\cite{cite5}.  This game got second-place in a coding contest at Matlab Central Exchange was an AI system that would play 2048 on its ~\cite{cite6}. We can consider a computer is playing this game against a human player which is the same as to think about it as a two player game. We can add some new ingredient, such as we can construct an algorithm for the computer which will make the human player moves more difficult. In this paper, we give a winning strategy and analyze the maximum value the player can reach for the standard two-dimensional board and also generalize for higher dimension. 

\medskip

When the human player (or the player) plays the game, initially the tiles of 2 and 4 appear randomly. We discuss the strategy that the computer uses in an attempt to defeat the player. After creating one tile, the player has many options to move the tile. We define $move$ $ by$ $ the$ $ player$ and for which the player will reach the maximum value. After one move, the computer also has many options to create a tile for which the next move for the player would be most difficult. First, the computer will create the tile by following its strategy. Then the player selects its move by winning strategy. By the winning strategy, the player can always reach the maximum value.

\subsection{Related Work}
In this connection, one may note that there are several works on the game 2048 but they are not related to our work. We have listed some of them here:
\begin{enumerate}
\item In the paper by Abdelkader et al.~\cite{cite7} has done in 2015, is about the complexity of a slightly adapted version and the proof of several natural decision problems turn out to be NP-Complete. They have also reduced from $\mathrm{3SAT}$ and implement their reduction as an online game, where we aim to develop a winning strategy for the game and its generalized version which is completely different from the study of complexity theory for 2048.
\item In the paper by Rahul Maheta~\cite{cite8} has done in 2014, which is about the analysis of the hardness of generating sequences of the moves, does not deal with what is the actual sequence of moves and how many of them are needed. 
\item In the paper by Bhargavi Goel~\cite{cite9} has done in 2017, which is about the analysis of the game for the standard two-dimensional board and the optimal strategy to ensure victory. She has been found a strategy to reach 2048 using the fuzzy theory which is completely probabilistic wherein our paper we have given a configuration the computer and a complete solution for the 2048 problem. We have generalized this problem also which is unique from other papers.
\item Note the papers,~\cite{cite10, cite11, cite12, cite13, cite14, cite16, cite19, cite20, cite21, cite22, cite23, cite24, cite25} on the game 2048 which are related to the statistical and mathematical analysis of the game 2048 where our aim is to find the winning strategy for 2048 and to generalize it.
\end{enumerate}

\subsection{Notations}
Here we summarize the symbols and notations that would be subsequently used throughout the paper.
\begin{center}
{\scriptsize
\begin{tabular}{|c||c|c|c|}
 \hline
 \multicolumn{2}{|c|}{List of Notations} \\ \hline
Notation & Name\\

\hline
$\mathcal{G}_{n_1 \times n_2 \times \ldots \times n_d}$ & $d$ dimensional board \\
\hline
$t$&   Tile for the board \\
\hline
$(i_t, j_t)$ & Position of a tile $t$ in 2D \\ 
\hline
$\left(i_{t,1}, i_{t,2}, i_{t,3}\ldots, i_{t,d}\right)$ & Position of a tile $t$ in $d$ dimensions \\
\hline
$Val(t)$& Value of a tile $t$  \\
\hline
$\mathcal{R}_t$   & Rectilinear neighbour of $t$ \\
\hline
${\mathcal{R}_{t}}^c = \mathcal{G}_{n_1 \times n_2 \times \ldots \times n_d} - \mathcal{R}_{t}$ &  Diagonal neighbour of $t$\\  
\hline
$({i_m}^s, {j_m}^s)$ & $m$-th term of the sequence $\mathcal{S}$ in 2D\\
\hline
$({i^s}_{{(m,1)}}, {i^s}_{{(m,2)}}, \ldots, {i^s}_{{(m,d)}})$ & $m$-th term of the sequence $\mathcal{S}$ \\
\hline
\end{tabular}}
\end{center}

\ \\

\section{2048 Problem for two-dimensional board $\mathcal{G}_{n_1\times n_2}$}\label{sec2}
In this section, we are going to construct the winning strategy for the player for the two-dimensional board $\mathcal{G}_{n_1 \times n_2}$ and also for the computer. To describe them mathematically, we need to define some terms for the board $\mathcal{G}_{n_1 \times n_2}$.

\begin{figure}[htbp]
\centering
\includegraphics[width=0.5\textwidth]{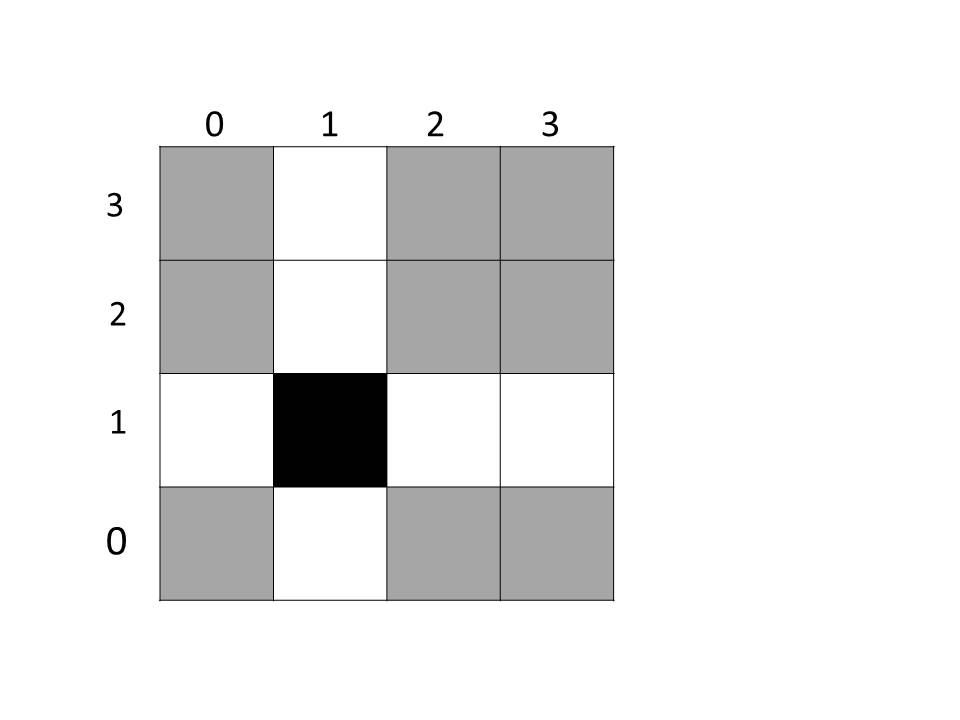}
\caption{The {white} shaded area is the Rectilinear neighbour $\mathcal{R}_t = (i_t, j_t)$ of the {black} coloured tile $t$ (here we have taken the board $\mathcal{G}_{4\times 4}$) }
\label{nearest}
\end{figure}

In the board $\mathcal{G}_{n_1 \times n_2}$, every tile position can be represented by a pair of co-ordinates $(i_t, j_t)$, where $0 \le i_t \le n_1-1$, and $0 \le j_t \le n_2-1$.
\begin{definition}
{\bf (Rectilinear neighbour)}: The rectilinear neighbour of a tile $(i_t, j_t)$,  is said to be the row $i_t$ and column $j_t$ of that tile $t$.
 If we consider the rectilinear neighbour of a tile $t$ as a set $\mathcal{R}_t$ then,
\begin{align*}
 \mathcal{R}_t =  & \{(i_t, *),(*,j_t), \text{ where the first coordinate * means} \\
& \text{ any integer lies between $[0, n_1-1]$ and the } \\
& \text{ second coordinate * means any integer lies } \\
& \text{ between $[0, n_2-1]$}\}.
\end{align*}

So we denoted a rectilinear neighbour of a tile $t$ by the set $\mathcal{R}_t$.
\end{definition}
In Figure~\ref{nearest}, we show the rectilinear neighbour $\mathcal{R}_t$, for the board $\mathcal{G}_{n_1 \times n_2}$.

\begin{figure}[htbp]
\centering
\includegraphics[width=0.5\textwidth]{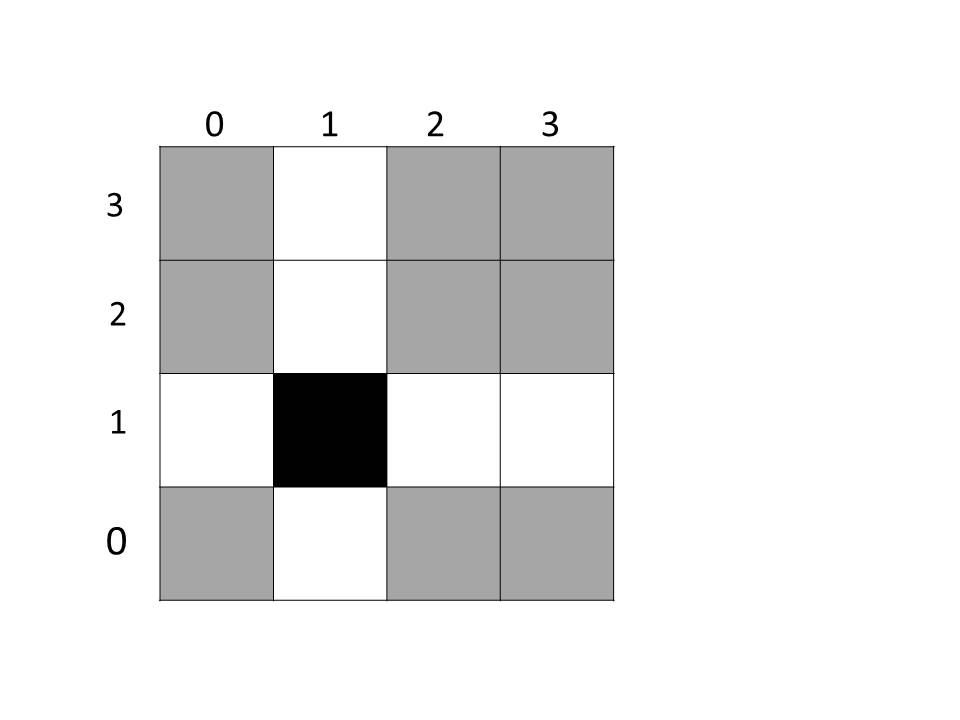}
\caption{The {gray} shaded area is the Diagonal neighbour ${\mathcal{R}_t}^c = \mathcal{G}_{n_1\times n_2}-{\mathcal{R}_t}$ of the {black} coloured tile $t$ (here we have taken the board $\mathcal{G}_{4\times 4}$)}
\label{diagonal}
\end{figure}
\begin{figure}[htbp]
\centering
\includegraphics[width=0.5\textwidth]{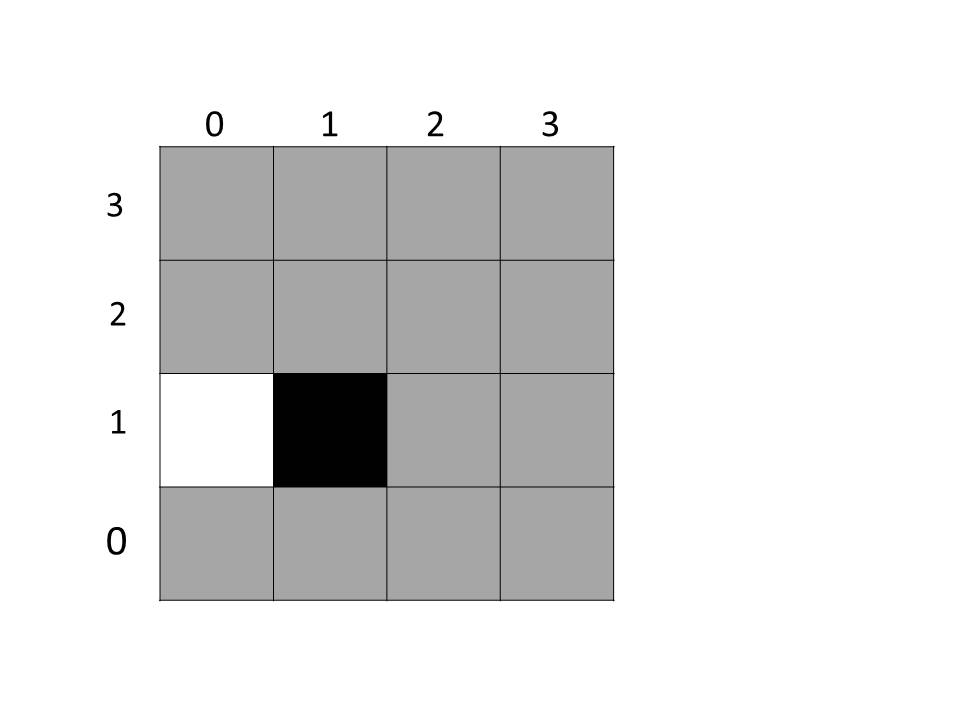}
\caption{The {black} coloured tile is the Previous tile $prev{(i_t,j_t)}$ of the {white} coloured tile $t$ (here we have taken the board $\mathcal{G}_{4\times 4}$) }
\label{previous}
\end{figure}

\begin{definition}
{\bf (Diagonal neighbour):} The diagonal neighbour of a tile $t$ is said to be any position that is not the rectilinear neighbour of that tile $t$.
So, the set $\mathcal{G}_{n_1\times n_2} - \mathcal{R}_t$ is the diagonal neighbour of the tile $t$, that is the complement set of the rectilinear neighbour where the universal set is the whole board.
We denoted a diagonal neighbour of a tile $t$ by the set ${\mathcal{R}_t}^c = \mathcal{G}_{n_1\times n_2} - \mathcal{R}_t$.
\end{definition}

In Figure~\ref{diagonal}, we show the diagonal neighbour for the board $\mathcal{G}_{n_1 \times n_2}$.

\medskip

[{\bf Note:} In this paper, we use the entire board as the $n$ neighbourhood region at a tile. In general, the neighbourhood region of a tile can be restricted to a certain radius.]

\begin{rem}
The set ${\mathcal{R}_t}^c = \mathcal{G}_{n_1\times n_2} - \mathcal{R}_t$ is the diagonal neighbour of the tile $t$, that is the complement set of rectilinear neighbour where the universal set is the whole board $\mathcal{G}_{n_1\times n_2}$. 
\end{rem}

\begin{definition}
{\bf (Previous tile):} For a tile position $t$, we define the previous tile of $(i_t, j_t)$, as a function of $i_t$ and $j_t$ as follows,

$prev{(i_t,j_t)}$
\begin{equation}
= \left\{ \begin{array}{ll}
    (i_t,j_t-1), & \mbox{ if $(n_1-i_t)$ mod 2 = 1 and $j_t$ = $n_2-1$};\\  & Or, \mbox{ $(n_1-i_t)$ mod 2 = 0 and $j_t$ = 0}; \\
    (i_t,j_t+1), & \mbox{ if $(n_1-i_t)$ mod 2 = 1 and $j_t$ $>$ $0$ }; \\
    (i_t,j_t-1), & \mbox{ if $(n_1-i_t)$ mod 2 = 0 and $j_t$ $<$ $n_2-1$ };
\end{array} \right.
\end{equation}
\end{definition}
In Figure~\ref{previous} we show the previous tile $prev{(i_t,j_t)}$ for the board $\mathcal{G}_{n_1\times n_2}$.

\begin{definition}
{\bf (Favourable situation):} If the value of the tile position ($i_t, j_t$) (of the tile $t$) is $2^m$ and the value of the previous tile $prev{(i_t,j_t)}$ is $2^{m+1}$ (where $1 \le m \le n_1 n_2$) then it is called the favourable situation. \end{definition}
In Figure~\ref{favourable}, we show the Favourable situation for the board $\mathcal{G}_{4 \times 4}$.
\begin{figure}[htbp]
\label{favourable}
\begin{center}
\def\mca#1{\multicolumn{1}{c}{#1}}
\def\mcb#1{\multicolumn{1}{c|}{#1}}
\renewcommand{\arraystretch}{2}
\begin{tabular}{c|c|c|c|c|c|}
  \mca{}  & \mca0 & \mca1 & \mca2 & \mca3 \\\cline{2-5}
  \mcb3   & $2^2$     & $2^3$    & $2^4$ & $2^5$     \\\cline{2-5}
  \mcb2   & $2^9$    & $2^8$   & $2^7$ & $2^6$   \\\cline{2-5}
  \mcb1 & $2^{10}$    & $2^{11}$    & $2^{12}$ & $2^{13}$    \\\cline{2-5}
  \mcb0 & $2^{17}$    & $2^{16}$     & $2^{15}$ & $2^{14}$     \\\cline{2-5}
\end{tabular}
\caption{Favourable situation for $\mathcal{G}_{4 \times 4}$ board}
\end{center}
\end{figure}

\begin{definition}\label{maxval}
{\bf (Maximum value):} Maximum value is the maximum value the player can achieve overall possible scenarios. 
\end{definition}

\begin{figure}[htbp]
\centering
\includegraphics[width=0.5\textwidth]{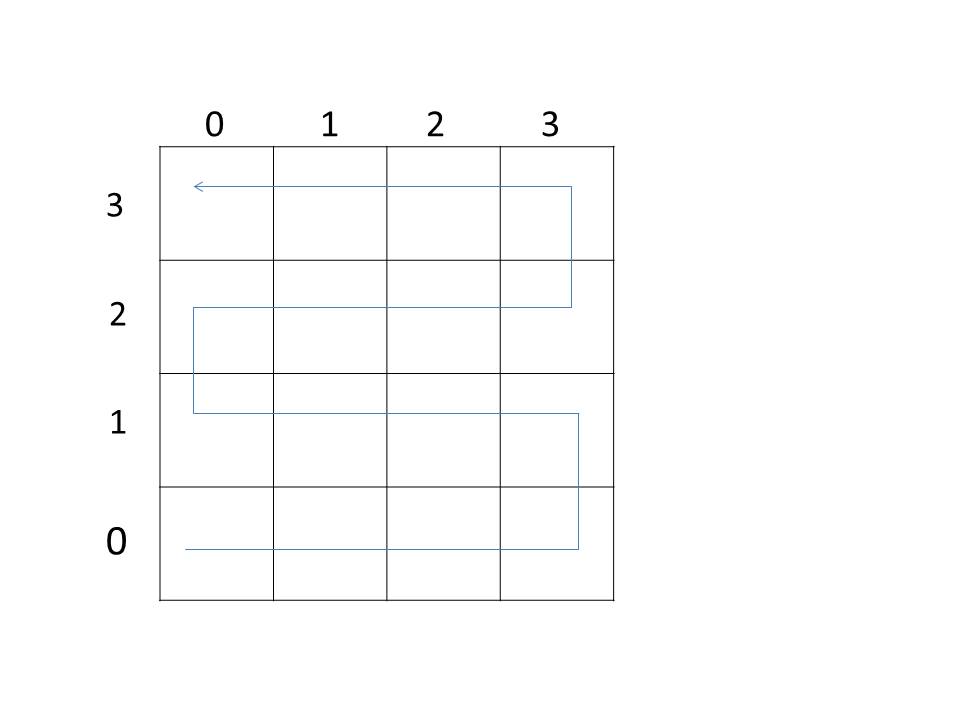}
\caption{Winning strategy for the player}
\label{winstat}
\end{figure}

\begin{definition}\label{target}
{\bf (Target cell):} Consider the tile $t_{max}$ currently at position $(i_{max} , j_{max})$ containing the maximum value say $2^l$. In order to increase the value at $(i_{max} , j_{max})$ we need to create second tile $t' $ of value $2^l$ at position $(i_{t'} , j_{t'})$, where $(i_{t'} , j_{t'})$ is the previous element of $(i_{max} , j_{max})$ in the sequence specified by $\mathcal{S}$ whose $m$-th term (for $m\geq 0$) is given by $({i_m}^s,{j_m}^s)$ where 
\begin{align}\label{eqtarget}
&{i_m}^s=q \text{ and } \\ \nonumber
&{j_m}^s={n_2}^{q\mod 2}-1+(-1)^{q\mod 2}r,
\end{align}
where $q=m/{n_2}$ (integer division) and $r=m\mod n_2$. 
\end{definition}
For example if $n_1=n_2=4$, then the sequence is given by the path shown in the figure~\ref{winstat}.

\begin{definition}
{\bf (Tile value):} Value of a tile $t$ (or tile value of $t$) is the numerical value\footnote{Clearly, which is in terms of $2^n$ for all $n\geq1$. The maximum value of $n$ depends on the size of the board.} of the tile $t$. It is denoted by $Val(t)$.
\end{definition}

\begin{definition}
{\bf (Merge):} Add up of the tiles $t$ and $t^{\prime}$ is called merge where $Val(t)$ = $Val(t^{\prime})$ for some tile $t$ and $t^{\prime}$.
\end{definition}

\begin{definition}
{\bf (Sub-board):} Consider two boards $\mathcal{G}_{n_1\times n_2}$ and $\mathcal{G}_{{n_1}^{\prime}\times {n_2}^{\prime}}$, with the inequality,
\begin{align*}
{n_1}^{\prime} \leq n_1 \text{ and } {n_2}^{\prime} \leq n_2.
\end{align*}
Then the board $\mathcal{G}_{{n_1}^{\prime}\times {n_2}^{\prime}}$ fits inside the board $\mathcal{G}_{n_1\times n_2}$. Then $\mathcal{G}_{{n_1}^{\prime}\times {n_2}^{\prime}}$ is called the sub-board of $\mathcal{G}_{n_1\times n_2}$. When the equality holds then sub-board itself is the entire board.
\end{definition}

For example a $2\times2$ board is always a sub-board of all $n\times n$ board with $n\geq2$.

\medskip

The game 2048 has been designed as a two player game. One player is the computer and the other one is the human player. In the original game, the computer creates a tile randomly anywhere in the board. So, it is clear that the tile appearance position is completely probabilistic. Sometimes some particular tile appearance helps the player to make the move easier. Here we have described a particular strategy for the computer for which the game will be more difficult for the player. The motivation of such design is to make the game harder for the player. Also, it helps to develop the winning strategy for the player because the probabilistic tile creation it is not possible to develop an algorithm for which the player will always win. This strategy for the computer which we are going to describe follows the rules of tile creation of the original game. Moreover, by the original design of the game 2048, the first move will be made by the computer that is the game will be started by the computer.

\medskip

{\bf Strategy for the Computer:} The computer will start the game by creating a tile in the empty board. We are going to describe its tile creating strategy here. \\

{\bf Step 1:} Computer will make its first move by creating a tile $t$ at a random position of the board $(t_t, j_t)$ where  $0 \le i_t \le n_1-1$ and $0 \le j_t \le n_2-1$. \\

{\bf Step 2:} Computer will create a tile $t^{\prime}$ with $Val(t^{\prime})=4$ (i.e., 4 valued tile) at the rectilinear neighbour $\mathcal{R}_t$ of  the tile position $(i_t, j_t)$ of the tile $t$, iff the value of the tile $t$ is greater than or equal to 4 i.e., $Val(t) \geq 4$. 
Otherwise, the computer will create a tile $t^{\prime}$ of value 2 (i.e., $Val(t^{\prime}) = 2$) in the diagonal neighbour ${\mathcal{R}_t}^c$ of the tile position $(i_t, j_t)$ for the tile $t$. \\

{\bf Step 3:} Computer will create a tile $t$ of value 2  (i.e., $Val(t) = 2$) at the position $(i_t, j_t)$, when there is a tile $t^{\prime}$ of value 2 (i.e., $Val(t^{\prime}) = 2$) in the rectilinear neighbour $\mathcal{R}_t$ of the tile position $(i_t, j_t)$. \\

 The computer will create a tile $t^{\prime}$ in the opposite direction of motion concerning the tile $t$ moved by the player.

\medskip

Now we write the winning strategy for the player. \\


\medskip
{\bf Winning strategy for the Player:} After creating the first tile by the computer the player will make its first move. Here we have described the winning strategy for the player. In the beginning of the game, there is only one 4-valued tile $t_1$ in the board, say, placed at a random position ($i_{t_1}, j_{t_1}$), where $0 \leq i_{t_1} \leq n_1-1$, $0 \leq j_{t_1} \leq n_2-1$. \\

{\bf Step 1:} The player will choose the value of $m=0$ to select the first target cell. For the value of $m=0$, the first term of the sequence $\mathcal{S}$ or the first target cell is as follows:
\begin{align*}
&q=m/{n_2} = 0/{n_2} = 0, \text{ then } {i^s}_{0}= q = 0. \\
&\text{ Since } r=0\pmod{n_2}, \text{ so } \\
& {j_0}^s = {n_2}^{0\pmod{2}}-1+(-1)^{0\pmod{2}}0 = 1-1+0=0.
\end{align*}
So, the first target cell is $(0, 0)$. The player will bring the tile $t_1$ to the position ($0, 0$), unless it is already in this position. If the player can bring the tile $t_1$ by one move then the player will do so. Otherwise, the player will bring $t_1$ to the rectilinear position of ($0, 0$) i.e., $\mathcal{R}_{(0, 0)}$. \\

For this move by the player, the computer will create another tile $t_2$ against the player on the board. Then the player will move the tile $t_1$ to the position ($0, 0$). As per Definition~\ref{def1}, $t_2$ will appear in the opposite direction of motion of $t_1$ and hence will not obstruct its movement to the position ($0, 0$). \\

{\bf Step 2:} In the second step the player will fix the target cell $t_{max}$ according to the sequence\footnote{the value of $m$ will varies over $0, 1, 2, \ldots, n_1n_2-1$.} $S_m$. Then the player will look at the value of the previous tile $\left(\text{ i.e., } prev\left(i_{t_{max}}, j_{t_{max}}\right)\right)$ of the target cell $t_{max}$. \\

{\bf Step 3:} In the next step, the player will try to make the same valued tile in the target cell $t_{max}$. For each tile, there are 4 possible moves for the board $\mathcal{G}_{n_1\times n_2}$ and for each move, there are 3 possible cases.

\smallskip

\begin{case}
If there is a $t$ with $Val(t) = 4$ in the rectilinear position $\mathcal{R}_{t_{max}}$ of the target cell $t_{max}$, then the player will bring that tile $t$ (where $Val(t) = 4$) in the target cell $t_{max}$ by one move.
\end{case}

\begin{case}
If there is a tile $t$  with $Val(t) = 2$ in the rectilinear position $\mathcal{R}_{t_{max}}$ of the target cell $t_{max}$, and there are two another tiles $t_1$ and $t_2$ with $Val(t_1) = Val(t_2) = 2$ in the board $\mathcal{G}_{n_1\times n_2}$ which can be merged, then the player will merge that tile $t_1$ and $t_2$ to get another tile $t_3$ where, $Val(t_3)$ = 4. Otherwise, the player will bring the tile $t$ (with $Val(t) = 2$) to the position $t_{max}$ by one move.
\end{case}

\begin{case}
If there is no tile $t$ in the rectilinear position $\mathcal{R}_{t_{max}}$ of the target cell $t_{max}$, then the player will bring a tile $t_1$ with $Val(t_1) = 4$ to the rectilinear position $\mathcal{R}_{t_{max}}$ of the target cell $t_{max}$ from the diagonal position ${\mathcal{R}_{t_{max}}}^c$ of the target cell $t_{max}$. To bring this tile to the rectilinear position $\mathcal{R}_{t_{max}}$ of $t_{max}$ the player will choose row wise move. If there is no tile $t$ with $Val(t) = 4$ in the board $\mathcal{G}_{n_1\times n_2}$ then the player will merge the tile $t_1$ and $t_2$ (with $Val(t_1) = Val(t_2) = 2$ )to get another tile $t_3$ where, $Val(t_3) = 4$.

\smallskip

When the player has made a tile $t$ with $Val(t) = 8$ in the target cell $t_{max}$ then according to the sequence $S_m$, the player will choose another target cell and will continue this process. When the player has made the same valued tile concerning the previous tile of the target cell $t_{max}$, then the player will merge those two tiles.
\end{case}

If the player will continue this game in this way, the player can make the favourable situation and the maximum value.

\begin{figure}[htbp]
\centering
\includegraphics[width=0.5\textwidth]{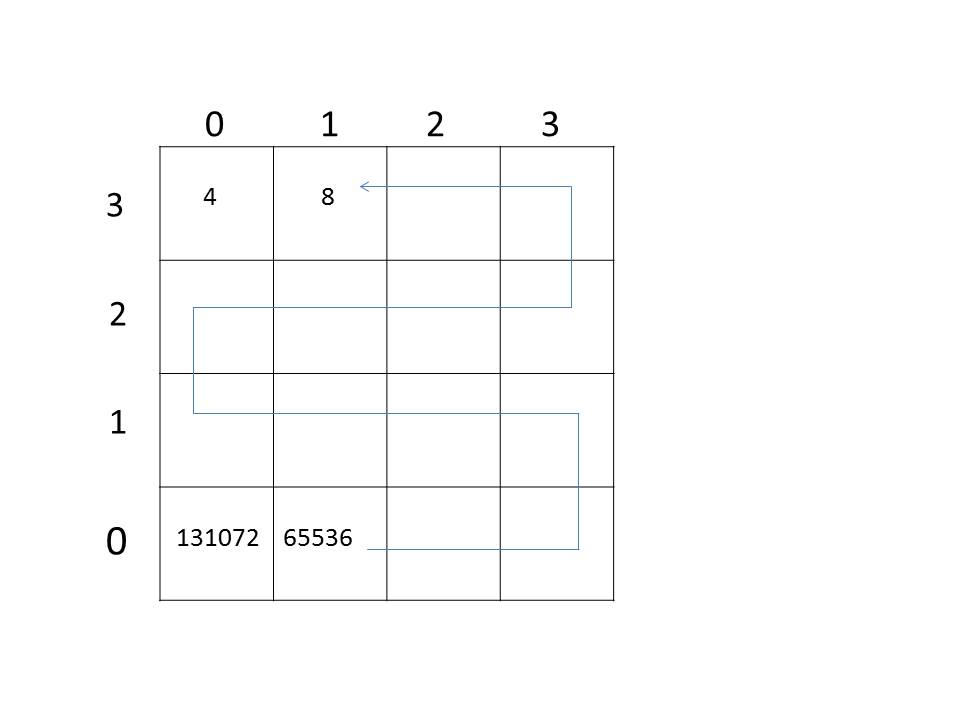}
\caption{Maximum value the player can reach}
\label{maxmin}
\end{figure}

\medskip

The described winning strategy is the algorithm for the player for which the player will always reach the maximum value. To reach the maximum value the player needs to make the favourable situation first. In the next result, we state and prove that if the player follows this winning strategy we have described above then it can make the favourable situation in the board.

\begin{lemma}\label{lem1}
The player can reach the favourable situation if it follows the winning strategy.
\end{lemma}

\pf The game will be started by the computer by creating a tile $t_1$ at the random position $(i_{t_1}, j_{t_1})$ for the board $\mathcal{G}_{n_1\times n_2}$. By the winning strategy of the player it will choose the first target cell (see Definition~\ref{target}) $t_{max}$ according to the sequence $\mathcal{S}$. We have defined the sequence $\mathcal{S}$ in Equation~\ref{eqtarget}. \\

According to that the player will choose its first target cell with the value $m=0$. Putting this value in Equation~\ref{eqtarget} we get,

\begin{align*}
&q=m/{n_2} = 0/{n_2} = 0 \text{ then } {i_0}^s= q = 0. \\
&r=0\pmod{n_2} \implies r=0 \text{ then } \\
& {j_0}^s = {n_2}^{0\pmod{2}}-1+(-1)^{0\pmod{2}}0 = 1-1+0=0.
\end{align*}

So the first target cell is $(0, 0)$. The player will bring the tile $t_1$ at the position $(0, 0)$. Then if we follow the winning strategy and the sequence the next target cell the player fixes with $m=1$. If we do the same calculation as the above equation we get that the next target cell is $(0, 1)$. Then the player will make another 4-valued tile $t_2$ at the position $(0, 1)$. There is a possible merge. So, the player will merge these two tiles and get another tile $t_3$ at the position $(0, 0)$ with $Val(t_3) = 8$. Likewise, again the player will make a 4-valued tile at $(0, 1)$ and another 4-valued tile at $(0, 2)$. Then after merging them, the player will get another 8-valued tile at $(0, 1)$. Already there was an 8-valued tile at the position $(0, 0)$. Then the player will merge them to get a tile with value 16 at the position $(0, 0)$. 

\medskip

The player will continue this process like this by choosing the value of $m$ (which varies over $0, 1, 2, \dots, n_1n_2-1$ for the board $\mathcal{G}_{n_1\times n_2}$) and fixing the target cell accordingly. If the player goes through this process then it is clear from the above argument that the player can make the favourable situation in the board $\mathcal{G}_{n_1\times n_2}$. Hence we have proved the following lemma. 
(A picture of the favourable situation is drawn in Figure~\ref{favourable}). \qed 

\medskip

It is clear that if the player follows the winning strategy then it can reach the favourable situation. Now we have to prove that if the player can make the favourable situation for the board $\mathcal{G}_{n_1\times n_2}$ then the player can reach the maximum value. In the next lemma, we prove this result for the board $\mathcal{G}_{2\times2}$.
\begin{lemma}\label{lem2}
For the board $\mathcal{G}_{2\times2}$ the maximum value the player can reach is $2^5=32$.
\end{lemma}

\begin{figure}[htbp]
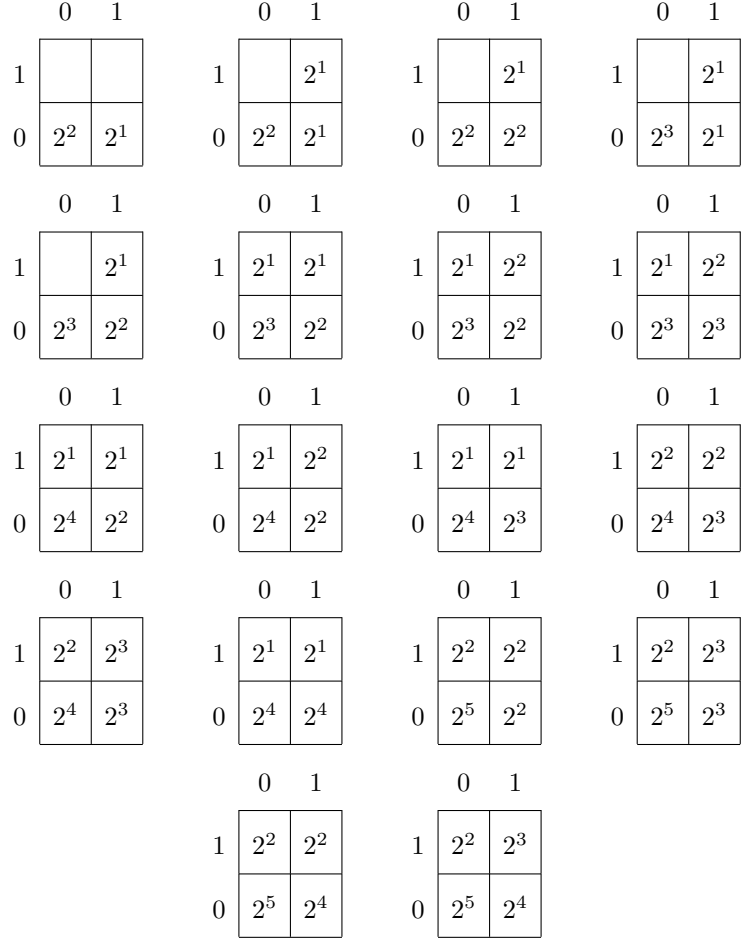

\begin{center}
\def\mca#1{\multicolumn{1}{c}{#1}}
\def\mcb#1{\multicolumn{1}{c|}{#1}}
\renewcommand{\arraystretch}{2}
\begin{minipage}{.2\textwidth}
\begin{tabular}{c|c|c|}
  \mca{}  & \mca0 & \mca1 \\\cline{2-3}
  \mcb1   &      &          \\\cline{2-3}
  \mcb0   & $2^2$    & $2^1$      \\\cline{2-3}
\end{tabular}
\end{minipage}
\begin{minipage}{.2\textwidth}
\begin{tabular}{c|c|c|}
  \mca{}  & \mca0 & \mca1 \\\cline{2-3}
  \mcb1   &      &   $2^1$       \\\cline{2-3}
  \mcb0   & $2^2$    & $2^1$      \\\cline{2-3}
\end{tabular}
\end{minipage}
\begin{minipage}{.2\textwidth}
\begin{tabular}{c|c|c|}
  \mca{}  & \mca0 & \mca1 \\\cline{2-3}
  \mcb1   &      &   $2^1$       \\\cline{2-3}
  \mcb0   & $2^2$    & $2^2$      \\\cline{2-3}
\end{tabular}
\end{minipage}
\begin{minipage}{.2\textwidth}
\begin{tabular}{c|c|c|}
  \mca{}  & \mca0 & \mca1 \\\cline{2-3}
  \mcb1   &      &   $2^1$       \\\cline{2-3}
  \mcb0   & $2^3$    & $2^1$      \\\cline{2-3}
\end{tabular}
\end{minipage}
\begin{minipage}{.2\textwidth}
\begin{tabular}{c|c|c|}
  \mca{}  & \mca0 & \mca1 \\\cline{2-3}
  \mcb1   &     &   $2^1$       \\\cline{2-3}
  \mcb0   & $2^3$    & $2^2$      \\\cline{2-3}
\end{tabular}
\end{minipage}
\begin{minipage}{.2\textwidth}
\begin{tabular}{c|c|c|}
  \mca{}  & \mca0 & \mca1 \\\cline{2-3}
  \mcb1   &  $2^1$    &   $2^1$       \\\cline{2-3}
  \mcb0   & $2^3$    & $2^2$      \\\cline{2-3}
\end{tabular}
\end{minipage}
\begin{minipage}{.2\textwidth}
\begin{tabular}{c|c|c|}
  \mca{}  & \mca0 & \mca1 \\\cline{2-3}
  \mcb1   &  $2^1$    &   $2^2$       \\\cline{2-3}
  \mcb0   & $2^3$    & $2^2$      \\\cline{2-3}
\end{tabular}
\end{minipage}
\begin{minipage}{.2\textwidth}
\begin{tabular}{c|c|c|}
  \mca{}  & \mca0 & \mca1 \\\cline{2-3}
  \mcb1   &  $2^1$    &   $2^2$       \\\cline{2-3}
  \mcb0   & $2^3$    & $2^3$      \\\cline{2-3}
\end{tabular}
\end{minipage}
\begin{minipage}{.2\textwidth}
\begin{tabular}{c|c|c|}
  \mca{}  & \mca0 & \mca1 \\\cline{2-3}
  \mcb1   &  $2^1$    &   $2^1$       \\\cline{2-3}
  \mcb0   & $2^4$    & $2^2$      \\\cline{2-3}
\end{tabular}
\end{minipage}
\begin{minipage}{.2\textwidth}
\begin{tabular}{c|c|c|}
  \mca{}  & \mca0 & \mca1 \\\cline{2-3}
  \mcb1   &  $2^1$    &   $2^2$       \\\cline{2-3}
  \mcb0   & $2^4$    & $2^2$      \\\cline{2-3}
\end{tabular}
\end{minipage}
\begin{minipage}{.2\textwidth}
\begin{tabular}{c|c|c|}
  \mca{}  & \mca0 & \mca1 \\\cline{2-3}
  \mcb1   &  $2^1$    &   $2^1$       \\\cline{2-3}
  \mcb0   & $2^4$    & $2^3$      \\\cline{2-3}
\end{tabular}
\end{minipage}
\begin{minipage}{.2\textwidth}
\begin{tabular}{c|c|c|}
  \mca{}  & \mca0 & \mca1 \\\cline{2-3}
  \mcb1   &  $2^2$    &   $2^2$       \\\cline{2-3}
  \mcb0   & $2^4$    & $2^3$      \\\cline{2-3}
\end{tabular}
\end{minipage}
\begin{minipage}{.2\textwidth}
\begin{tabular}{c|c|c|}
  \mca{}  & \mca0 & \mca1 \\\cline{2-3}
  \mcb1   &  $2^2$    &   $2^3$       \\\cline{2-3}
  \mcb0   & $2^4$    & $2^3$      \\\cline{2-3}
\end{tabular}
\end{minipage}
\begin{minipage}{.2\textwidth}
\begin{tabular}{c|c|c|}
  \mca{}  & \mca0 & \mca1 \\\cline{2-3}
  \mcb1   &  $2^1$    &   $2^1$       \\\cline{2-3}
  \mcb0   & $2^4$    & $2^4$      \\\cline{2-3}
\end{tabular}
\end{minipage}
\begin{minipage}{.2\textwidth}
\begin{tabular}{c|c|c|}
  \mca{}  & \mca0 & \mca1 \\\cline{2-3}
  \mcb1   &  $2^2$    &   $2^2$       \\\cline{2-3}
  \mcb0   & $2^5$    & $2^2$      \\\cline{2-3}
\end{tabular}
\end{minipage}
\begin{minipage}{.2\textwidth}
\begin{tabular}{c|c|c|}
  \mca{}  & \mca0 & \mca1 \\\cline{2-3}
  \mcb1   &  $2^2$    &   $2^3$       \\\cline{2-3}
  \mcb0   & $2^5$    & $2^3$      \\\cline{2-3}
\end{tabular}
\end{minipage}
\begin{minipage}{.2\textwidth}
\begin{tabular}{c|c|c|}
  \mca{}  & \mca0 & \mca1 \\\cline{2-3}
  \mcb1   &  $2^2$    &   $2^2$       \\\cline{2-3}
  \mcb0   & $2^5$    & $2^4$      \\\cline{2-3}
\end{tabular}
\end{minipage}
\begin{minipage}{.2\textwidth}
\begin{tabular}{c|c|c|}
  \mca{}  & \mca0 & \mca1 \\\cline{2-3}
  \mcb1   &  $2^2$    &   $2^3$       \\\cline{2-3}
  \mcb0   & $2^5$    & $2^4$      \\\cline{2-3}
\end{tabular}
\end{minipage}
\caption{The game for the board $\mathcal{G}_{2\times2}$}
\label{example}
\end{center}
\end{figure}

\pf The board $\mathcal{G}_{2\times2}$ is basically a $2\times2$ board (see Figure~\ref{example}). There are $2\times2=4$ boxes. It is clear from the pigeonhole principle~\cite{cite26} that we can put a maximum of 4 tiles inside the board because in each box we can put only one tile. So, from this, it is clear that the maximum possible value the player can reach is $2^4=16$. Let us play the game following the described winning strategy, then we can make the favourable situation in the board $\mathcal{G}_{2\times2}$ and from Lemma~\ref{lem1} it follows that we can reach the maximum value. 

First, the computer will create a tile of value 4 at the random position of the board. The player will fix the target cell and bring it to $(0, 0)$. Then make a 4 valued tile at $(0, 1)$ and a 2 valued tile at $(1, 1)$. In this situation, the computer will create a tile at $(0, 1)$ of value 2 (follows from the strategy of the computer). Then the player will merge them and get a 8 valued tile at $(0, 0)$ and a 4 valued tile at $(0, 1)$ and another 4 valued tile at $(1, 1)$ . Again the player will merge and get a 16 valued tile at position $(0, 0)$ (as drawn in Figure~\ref{example}). 

Now see the Figure~\ref{example}, after making the tile with value 16 at position $(0, 0)$ and the player can make the favourable situation in the board and if a 4-valued tile appears at the position $(1, 0)$, then the player can make $2^5=32$, which is our desired maximum value. So for the board, $\mathcal{G}_{2\times2}$ the maximum value the player can reach is $2^5=32$. 
\qed 

\medskip

One can ask why a 4-valued tile will appear at position $(1, 0)$ after reaching the favourable situation in the board. Well, if we see the strategy for the computer and Definition~\ref{maxval}, then it says that the maximum value the player can reach over all possible scenarios. Appearing a 4-valued tile at the position $(1, 0)$ is a possible scenario and appearing a 2-valued tile at the same position is also a possible scenario, from the strategy of the computer the best case that is the first possible scenario will happen and the player can reach the maximum value.

Well, what we have shown in the Lemma~\ref{lem2} is for a particular board. Now we generalize and prove this statement for any $n_1\times n_2$ board $\mathcal{G}_{n_1\times n_2}$. To prove this result, we construct the next theorem.

\begin{theorem}
\label{twodasym}
 In the board $\mathcal{G}_{n_1 \times n_2}$, the player can reach the maximum value $2^{n_1 \times n_2 +1}$  $($where $n_1 \ge 1, n_2 \ge 1)$ . 
\end{theorem}

{\bf Discussion:} There are $n_1\times n_2$ many boxes in the board $\mathcal{G}_{n_1\times n_2}$ and one tile can be placed in each boxes. From Lemma~\ref{lem2} and from the pigeonhole principle~\cite{cite26} it is clear that there are $n_1\times n_2$ many boxes in the board $\mathcal{G}_{n_1\times n_2}$ and the maximum value the player can reach is $2^{n_1\times n_2}$. But as we have discussed before that according to the strategy of computer and by Definition~\ref{maxval}, at the position $(n_1-1, 0)$ a 4-valued tile will appear after the player reaches the favourable situation playing through the winning strategy. then the player can make the maximum value $2^{n_1\times n_2+1}$ for the board $\mathcal{G}_{n_1\times n_2}$. This discussion is a combinatorial argument for this theorem. Now we prove the theorem using the method of mathematical induction.

\medskip

\pf We would like to use inductive argument on $n_1$ and $n_2$ respectively to prove this result. First, we use an induction argument on $n_1$, which is the first step of the proof.

\medskip

{\bf Step 1:} For the base case, take $n_1$ = 2 and $n_2$ = 2, the player can reach the maximum value,
\begin{align*}
2^{(2\times2) +1} = 2^5 = 32,
\end{align*}
which follows from Lemma~\ref{lem2}.

\medskip

Let us assume that for the board $\mathcal{G}_{n_1 \times n_2}$, the player can reach the maximum value $2^{(n_1 n_2+1)}$.

\medskip

Now for the board $\mathcal{G}_{(n_1+1) \times n_2}$, there are 
\begin{align*}
(n_1+1)n_2 = n_1n_2+n_2, 
\end{align*}
many boxes.

The player is playing the game through the winning strategy, from the assumption of the inductive argument  the player can reach $2^{(n_1n_2+1)}$, because the board $\mathcal{G}_{n_1\times n_2}$ is a sub-board of the board $\mathcal{G}_{(n_1+1) \times n_2}$. Then,

\begin{align*}
(n_1+1)n_2 - n_1n_2 = n_1n_2 + n_2 - n_1n_2 = n_2,  
\end{align*}

i.e., $n_2$ many cells are left in the board $\mathcal{G}_{(n_1+1) \times n_2}$. So, the player can make the favourable situation. After fixing the target cell $t_{max}$ the player will try to make the same valued tile at the position of the tile $t_{max}$ which is $(i_{t_{max}}, j_{t{max}})$ with respect to it's previous tile $prev{\left(i_{t_{max}},j_{t_{max}}\right)}$. If the value of $t_{max}$ is equal to it's previous tile $prev{(i_t,j_t)}$ the player will merge them. The player is playing the game using the winning strategy and it is proved in the Lemma~\ref{lem1} that the player can make the favourable situation in the board. To get the favourable situation, the player will make same value tile with respect to it's previous tile $prev{(i_t,j_t)}$ and easily the player can reach, 

\begin{align*}
2^{n_1n_2+1+1} = 2^{n_1n_2+2}.
\end{align*}

The player will continue this process for $n_2$ vacant cells to reach the maximum value.

\begin{align*}
 &=2^{n_1n_2+1} \times 2^{n_2}\\
&=2^{{(n_1n_2+n_2)}+1}\\
&=2^{{((n_1+1)n_2)}+1}. 
\end{align*}

Hence, we have proved the result for the induction on $n_1$. Now, we are going to use inductive argument on $n_2$, which is the second step of our proof.

\medskip

{\bf Step 2:} For the base case, take $n_1$ = 2 and $n_2$ = 2, the player can reach the maximum value,
\begin{align*}
2^{(2\times2) +1} = 2^5 = 32,
\end{align*}
which follows from Lemma~\ref{lem2}. The base is same as the first step.

\medskip

Let us assume that for the board $\mathcal{G}_{n_1 \times n_2}$, the player can reach the maximum value $2^{(n_1n_2+1)}$.

Now for the board $\mathcal{G}_{n_1 \times( n_2+1)}$ there are 

\begin{align*}
n_1(n_2+1) = n_1n_2+n_1, 
\end{align*}

many boxes. After reaching $2^{(n_1n_2+1)}$ there

\begin{align*}
(n_2+1)n_1 - n_1n_2 = n_1n_2 + n_1 - n_1n_2 = n_1,  
\end{align*}

i.e., $n_1$ vacant cells are left. For the similar inductive argument, we have used on $n_1$ (in step 1), the player can reach the maximum value

\begin{align*}
 &=2^{n_1n_2+1} \times 2^{n_1}\\
&=2^{{(n_1n_2+n_1)}+1}\\
&=2^{{(n_1(n_2+1)}+1}. 
\end{align*}

Hence, by the induction hypothesis the player can reach the maximum value $2^{(n_1n_2+1)}$ for the  two-dimensional board $\mathcal{G}_{n_1 \times n_2}$, as desired. \qed

\begin{corollary}
For the special case $n_1 = n_2 = n$, that is for the board $\mathcal{G}_{n\times n}$ the maximum value the player can reach is $2^{(n^2+1)}$. \end{corollary}

\pf By Theorem~\ref{twodasym}, for $n_1 = n_2 = n$ the maximum value the player can reach is $2^{(n^2+1)}$ for the board $\mathcal{G}_{n\times n}$. \qed

\subsection{Algorithm for the two-dimensional board $\mathcal{G}_{n_1 \times n_2}$}

In this section, we write the algorithm for the winning strategy of the player, which is the main result of our paper. Moreover, we have also written the algorithm for the strategy the computer will follow, this algorithm (for the computer)\footnote{We have written the algorithm for the computer for the convenience of the reader. To understand the winning strategy for the player it is also important to know the tile creating an algorithm of the computer because the player will make it moves according to the tile creation.} has been designed from the original 2048 game under certain conditions. First, let us write the algorithm for the computer.

Note that both of these algorithms in this section has been written for the board $\mathcal{G}_{n_1\times n_2}$.

\medskip

The algorithm we are going to write for the computer, for each step the computer will create a tile in the opposite direction of motion concerning the tile moved by the player.

\begin{algorithm}[htbp]
\ Create tile $t$ at $(i_t, j_t)$ with $0\leq i_t\leq n_1-1$ and $0\leq j_t\leq n_2-1$. \\
\ {\bf If} there exist a tile $t$ with $Val(t) \geq 4$ {\bf then} \\
\ Create tile $t^{\prime}$ with $Val(t^{\prime}) = 4$ in $\mathcal{R}_t$ \\
\ {\bf else,} create tile $t^{\prime}$ with $Val(t^{\prime}) = 4$ in ${\mathcal{R}_t}^c$ \\
\ {\bf If} there exist a tile $t^{\prime}$ with $Val(t^{\prime}) = 2$ in $\mathcal{R}_t$ {\bf then} \\
\ Create tile $t$, with $Val(t)=2$ \\ 
\ {\bf else,} create tile $t$ with $Val(t)=2$ randomly at empty cells $(i_t, j_t)$.
\caption{\textbf Algorithm for the strategy of the Computer for the board $\mathcal{G}_{n_1\times n_2}$}
\label{algomax1}
\end{algorithm}

After computer creates its first tile on the board now it's the turn of the player to make its first move. Next, we write the algorithm for the player for which the player will reach the maximum value of this game, which we have described in the previous section as the \textit{Winning strategy of the player}. Now let us consider that the computer has appeared its first tile $t_1$ (say) on the board.

\begin{algorithm}[htbp]
\SetAlgoLined
\KwResult{The Player will make the favourable situation to reach the maximum value}
 Choose the value of $m=0$ to fix the first target cell $t_{(1,max)}$ calculate $({i_{(1,max)}}^s, {j_{(1,max)}}^s) = (0, 0)$. \\
 Move tile $t_1$ to bring it at the position $(0, 0)$. \\
 
 \While{The value of $m$ varies over $0, 1, \ldots,n_1n_2-1$}{
  calculate the target cells ${i_m}^s=q$ \& ${j_m}^s={n_2}^{q\pmod{2}}-1+(-1)^{q\pmod{2}}r$, (where $q=m/{n_2}$ (integer division) and $r=m\pmod{n_2}$) respectively. \;
  \eIf{there exist tile $t$ with $Val(t)=4$ at $\mathcal{R}_{t_{max}}$}{
  bring tile $t$ at $(i_{t_{max}}, j_{t_{max}})$ by one move. \;
   }{
   go to the next step\;
  }
    
    \eIf{there exist tile $t$ with $Val(t)=4$ at $\mathcal{R}_{t_{max}}$ and $t_1, t_2$ with $Val(t_1)=Val(t_2)=2$}{
  Merge $t_1$ and $t_2$ to get $t_3$ with $Val(t_3)=4$. \;
   }{
   Bring tile $t$ with $Val(t) = 2$ at $(i_{t_{max}}, j_{t_{max}})$ by one move \;
  }
  
  \eIf{there exist no tile $t$ at $\mathcal{R}_{t_{max}}$}{
  Choose row wise move and bring $t_1$ with $Val(t_1)$ at $\mathcal{R}_{t_{max}}$ from ${\mathcal{R}_{t_{max}}}^{c}$. \;
   }{
   go to the next step \;
  }
  
    \eIf{there exist no tile $t$ with $Val(t) = 4$ in the board $\mathcal{G}_{n_1\times n_2}$}{
  Merge $t_1$ and $t_2$ with $Val(t_1)=Val(t_2)=2$ to get $t_3$ with $Val(t_3)=4$ \;
   }{
   Return to step 5 \;
  }

    }
 
\caption{\textbf{Algorithm for winning strategy of the Player for the board $\mathcal{G}_{n_1\times n_2}$}}
\label{algomax2}
\end{algorithm}
\section{Generalization of 2048 to higher dimensions}

In the previous section, we have analyzed the game 2048 for the two-dimensional board and developed an algorithm for the player to win the game always. Now we move to the generalized version of this game, that is the generalization of 2048 in higher dimensions. In this section, we are going to generalize the game 2048 in the most general form, i.e., for the board $\mathcal{G}_{n_1 \times n_2 \times \ldots \times n_{d-1} \times n_d}$ in $d$-dimensions. Likewise, also in this section, we develop a winning strategy for the player and also construct the algorithm. Using these we count the maximum value the player can reach in $d$-dimensions.

\medskip

Every $d$-dimensional tile position of the tile $t$ (of the  board $\mathcal{G}_{n_1 \times n_2 \times \ldots \times n_{d-1} \times n_d}$) will thought as $d$ tuples of coordinates $\left(i_{t,1},\ldots, i_{t,d}\right)$. Like the previous section we define some terms for the higher dimensions and the other standard definitions will also holds for the higher dimensional board also. Let us define some terms for the board $\mathcal{G}_{n_1 \times n_2 \times \ldots \times n_{d-1} \times n_d}$. 

\medskip

\begin{definition}
{\bf (Rectilinear neighbour in $d$ dimensions)}: Every tile position in $d$-dimensional board $\mathcal{G}_{n_1 \times n_2 \times \ldots \times n_{d-1} \times n_d}$ can be represented by the co-ordinates $\left(i_{t,1},\ldots, i_{t,d}\right)$ where $0 \le i_j \le n_j-1$, where $j$ = $1, 2, \ldots ,d$ . When the co-ordinates $i_j$ (for $j = 2, \ldots ,d$) are fixed and $i_1$ varies then there is a row along $n_{i_1}$. Similarly, when $i_j$ varies for the different values of $j = 1, 2, \ldots ,d$ there are rows along $i_j.$ These rows are called the rectilinear neighbour of the tile $\left(i_{t,1},\ldots, i_{t,d}\right)$ where $0 \le i_j \le n_j-1$.
 If we consider the rectilinear neighbour of a tile $t$ as a set $\mathcal{R}_t$ then,
\begin{align*} 
\mathcal{R}_t =& \{\left(i_{t,1},\ldots, i_{t,d}\right): \text{ where }0 \le i_j \le n_j-1, \text{where, } \\
&j = 1, 2, \ldots ,d\text{ for the $d$ tuples of coordinates }\}.
\end{align*}
So we denoted a rectilinear neighbour of a tile $t$ by the set $\mathcal{R}_t$.
\end{definition}

\begin{definition}
{\bf (Diagonal neighbour in $d$ dimensions)}: Diagonal neighbour of a tile $t$ (of the  board $\mathcal{G}_{n_1 \times n_2 \times \ldots \times n_{d-1} \times n_d}$) is said to be any position that is not the rectilinear neighbour of that tile  $t$.
So, the set $\mathcal{G}_{n_1 \times n_2 \times \ldots \times n_{d-1} \times n_d} - \mathcal{R}_t$ is the diagonal neighbour of the tile $t$, that is the complement set of the rectilinear neighbour where the universal set is the whole board $\mathcal{G}_{n_1 \times n_2 \times \ldots \times n_{d-1} \times n_d}$.
We denoted a diagonal neighbour of a tile $t$ by the set 
${\mathcal{R}_t}^c = \mathcal{G}_{n_1 \times n_2 \times \ldots \times n_{d-1} \times n_d} - \mathcal{R}_t$.
\end{definition}

\begin{definition}
\label{def1}
{\bf (Move of the player)}: The player chooses a tile $t$ for the board $\mathcal{G}_{n_1\times n_2}$, and the direction $\vec{r}$ in which to move the chosen tile. There are $2n$ options for directions; $n$ mutually orthogonal axis-parallel directions denoted by $\vec{a_1}, \vec{a_2}, \ldots, \vec{a_n}$ and their opposite directions, denoted by, $-{\vec{a_1}}, -{\vec{a_2}}, \ldots, -{\vec{a_n}}$. Once the player chooses that direction then excluding the direction $\vec{r}$ we can identify the $(n-1)$ dimensional hyper-plane $h$ containing the tile $t$. Once the player initiates the move, each tile in the hyper-plane $h$ moves furthest possible distance along the direction $\vec{d}$. 
\end{definition}

\medskip

In the previous section, we have discussed the strategy for the computer in the two-dimensional board. The computer will always start the game and for $d$-dimensions the computer follows the standard rules of the game 2048. Instead of creating a tile it will create a $d$-dimensional hypercube in the board. So, we do not need to describe the strategy for the computer in higher dimensions again. We can write the winning strategy for the player directly for the board $\mathcal{G}_{n_1 \times n_2 \times \ldots \times n_{d-1} \times n_d}$. 

\medskip

{\bf Winning strategy for the player in the board $ \mathcal{G}_{n_1 \times n_2 \times \ldots \times n_{d-1} \times n_d}$:} The player starts the game after the first tile created by the computer. We have described the winning strategy for the player for two dimensional board previously, here we are going to generalize that idea for higher dimensions. In the beginning of the game there is only a 4-valued hypercube $t_1$ in the board, say, placed at random position $((i_{t,1},\ldots, i_{t,d})$, where $0\leq i_{t,1}\leq n_1-1, 0\leq i_{t,2}\leq n_2-1 \ldots, 0\leq i_{t,d}\leq n_d-1$.

\medskip

{\bf Step 1:} As first step, the player fixes the direction $\vec{a_2}$. As we know, $||a_i|| \le n_i-1$ for all $i = 1, 2, \ldots ,d$ and $n_i \ge 1$,  when the player will reach $n_2-1$ along the direction $\vec{a_2}$, it will fix the direction $\vec{a_1}$ for one step.   

\medskip

{\bf Step 2:} In this step, the player goes through the direction $\vec{(-a_2)}$ and after reaching $((n_2-1)-(n_2-1)) = 0$, the player will fix the direction $\vec{a_1}$ again. After completing $\vec{a_1}$ and $\vec{a_2}$ in two dimensions the player will go to 3-dimension.  

\medskip

{\bf Step 3:} For 3-dimensional board $\mathcal{G}_{n_1\times n_2\times n_3}$, we can think it as $n_3$ many layers of $\mathcal{G}_{n_1\times n_2}$ board. In three dimensions, the player will fix the direction $\vec{a_3}$. After completing three dimensions the player will go to four dimensions and will fix the direction $\vec{a_4}$ and so on, because we considered a $(d+1)$-dimensional $\mathcal{G}_{n_1 \times n_2 \times \ldots \times n_d \times n_{d+1}}$ board as $n_{d+1}$ many $d$-dimensional $\mathcal{G}_{n_1 \times n_2 \times \ldots \times n_d}$ boards stacked up on one another. 
A $n_1\times n_2\times \ldots \times n_d$ $d$-dimensional grid = $n_d$ many $n_1\times n_2\times \ldots \times n_{d-1}$ dimensional grid. In the sequence, after completing one $d-1$ dimensional grid we make one step jump to the neighbouring cell of the next $d-1$ dimensional grid. 

\medskip




For each tile, there are $2n$ possible moves and for each move, there are 3 possible cases as we discussed in the winning strategy for the board $\mathcal{G}_{n_1\times n_2}$ (see section~\ref{sec2}).

\medskip

\begin{rem}
In two-dimensional board, the player has fixed the target cell according to the sequence $\mathcal{S}$ but when the player is on the higher dimensions then the player fixes the direction respectively.
\end{rem}

We have described the winning strategy for the player above now we prove that using this winning strategy the player can reach the favourable situation in the $d$-dimensional board. We state and prove this result in the next lemma.

\usetikzlibrary{positioning}
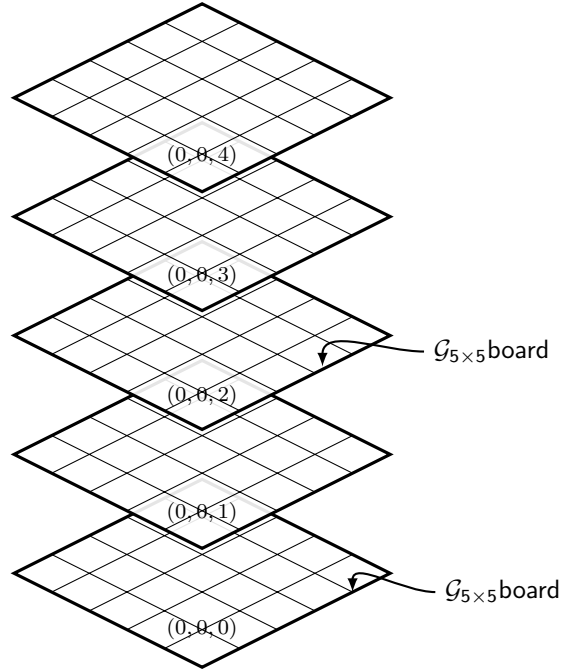
\begin{figure}
\begin{center}
\begin{tikzpicture}[scale=0.5,every node/.style={minimum size=1cm},on grid]
        
    \begin{scope}[
            yshift=0,every node/.append style={
            yslant=0.5,xslant=-1},yslant=0.5,xslant=-1
            ]
        \fill[white,fill opacity=2.0] (0,0) rectangle (5,5);
        \draw[step=10mm, black] (0,0) grid (5,5);
        \draw[black, very thick] (0,0) rectangle (5,5);
    \end{scope}
        
    \begin{scope}[
        yshift=90,every node/.append style={
            yslant=0.5,xslant=-1},yslant=0.5,xslant=-1
                     ]
        \fill[white,fill opacity=.9] (0,0) rectangle (5,5);
        \draw[step=10mm, black] (0,0) grid (5,5);
        \draw[black, very thick] (0,0) rectangle (5,5);;
    \end{scope}
        
    \begin{scope}[
        yshift=180,every node/.append style={
        yslant=0.5,xslant=-1},yslant=0.5,xslant=-1
                     ]
        \fill[white,fill opacity=.9] (0,0) rectangle (5,5);
        \draw[step=10mm, black] (0,0) grid (5,5);
        \draw[black, very thick] (0,0) rectangle (5,5);
    \end{scope}
        
    \begin{scope}[
        yshift=270,every node/.append style={
            yslant=0.5,xslant=-1},yslant=0.5,xslant=-1
          ]
        \fill[white,fill opacity=0.9] (0,0) rectangle (5,5);
        \draw[step=10mm, black] (0,0) grid (5,5);
         \draw[black, very thick] (0,0) rectangle (5,5);

    \end{scope}
        
    \begin{scope}[
        yshift=360,every node/.append style={
        yslant=0.5,xslant=-1},yslant=0.5,xslant=-1
                  ]
        \fill[white,fill opacity=.9] (0,0) rectangle (5,5);
        \draw[step=10mm, black] (0,0) grid (5,5);
        \draw[black, very thick] (0,0) rectangle (5,5);
    \end{scope} 

   \draw[-latex,thick] (6.2,2) node[right]{$\mathsf{\mathcal{G}_{5\times 5} board}$}
         to[out=180,in=90] (4,2);
\draw[-latex,thick](5.9,8.4)node[right]{$\mathsf{\mathcal{G}_{5\times 5} board}$}
        to[out=180,in=90] (3.2,8);

\fill[black,font=\footnotesize]
        (0,0) node [above] {$(0,0,0)$}
       (0,3.1) node [above] {$(0,0,1)$}
        (0,6.2) node [above] {$(0,0,2)$}
       (0,9.4) node [above] {$(0,0,3)$}    
       (0,12.6) node [above] {$(0,0,4)$};
        
\end{tikzpicture}
\caption{Winning strategy for $3$-dimensional board (here we have shown for $\mathcal{G}_{5\times5\times5}$ board)}
\label{wind}

\end{center}
\end{figure}

\medskip

\begin{lemma}
The player can reach the favourable situation if it follows the winning strategy for the $d$-dimensional board $\mathcal{G}_{n_1\times n_2\times \ldots \times n_d}$.
 \end{lemma}

\pf We have proved the same result for the two-dimensional board in Lemma~\ref{lem1}. The basic idea for the $d$-dimensions is also same but here instead of fixing the target cell the player fixes the direction. Let us enlightened this result for the $d$-dimensions. 

\medskip

The player will finish the two-dimensional game first then it will go to the three dimensions and after that four dimensions and so on. So the player has reached the favourable situation as well as the maximum value in the two-dimensional board. Now the player is moving to the higher dimension, so the player fixes the direction to move to the three-dimensional board. As we know that a three dimensional board $\mathcal{G}_{n_1\times n_2\times n_3}$ is $n_3$ many two dimensional board $\mathcal{G}_{n_1\times n_2}$ stacked upon each other. After fixing the direction the player will continue the game fixing the target cell according to the given sequence $\mathcal{S}$ and in this way, the player will complete the three-dimensional board. Then moves to four dimensions and so on. It is clear from Lemma~\ref{lem1} that the player will reach the favourable situation for the two-dimensional board, in the same way, it will reach the favourable situation in each of two-dimensional board in three dimensions. In this way, the player will reach the favourable situation for all $d$ dimensional board, as desired.

\qed 

\medskip

Now we state and prove that the maximum value the player can reach in the $d$ dimensional board $\mathcal{G}_{n_1 \times n_2 \times \ldots \times n_d}$.

\medskip

\begin{theorem}\label{thmd}
In the $d$-dimensional board $\mathcal{G}_{n_1 \times n_2 \times \ldots \times n_d}$, the player can reach the maximum value $2^{(\prod_{i=1}^{d}n_i) + 1}$.
\end{theorem}

\pf We will prove it using inductive argument on $d$. 

For any $d$ dimensional board $\mathcal{G}_{n_1 \times n_2 \times \ldots \times n_d}$, if we use inductive argument on $d$ then the base case of induction holds for $d=2$, i.e., the board $\mathcal{G}_{n_1 \times n_2}$. This is nothing but the Theorem~\ref{twodasym}, which we have already proved in the previous section. 

\medskip

Now let us assume that for $d$-dimensional board $\mathcal{G}_{n_1 \times n_2 \times \ldots \times n_d}$, the player can reach the maximum value $2^{(\prod_{i=1}^{d}n_i) + 1}$.

A $(d+1)$-dimensional board $\mathcal{G}_{n_1 \times n_2 \times \ldots \times n_d \times n_{d+1}}$ can be considered as $n_{d+1}$ many $d$-dimensional boards $\mathcal{G}_{n_1 \times n_2 \times \ldots \times n_d}$  (called sub-boards here) stacked up on one another.  

After reaching the maximum value in a $d$-dimensional sub-board, the player can continue the game in the adjacent $d$-dimensional sub-board.

By the winning strategy, the player can continue the game when at the $(0, 0, \ldots ,1)$-th position a $(d+1)$ dimensional hyper-cube of 8 is there and at the $(0, 0, \ldots ,0)$-th position a $(d+1)$ dimensional hyper-cube of 4 is there and at the $(1, 0, \ldots ,0)$-th position a  $(d+1)$ dimensional hyper-cube of bigger value of 4 and 8 is there, then the game will be end. So, the maximum value is reached by the player $2^{(\prod_{i=1}^{d+1}n_i) + 1}$.

Now, we use inductive argument on $n_i$'s, each dimension size.

First we take the board $\mathcal{G}_{n_1 \times n_2 \times \ldots \times n_d}$ in $d$ dimensional plane. The player can reach the maximum value $2^{(\prod_{i=1}^{d}n_i) + 1}$.

Now, for the board $\mathcal{G}_{(n_1 +1) \times n_2 \times \ldots \times n_d}$ in $d$-dimensional plane. After reaching $2^{(\prod_{i=1}^{d}n_i) + 1}$ there are $n_1 n_2 \ldots n_d$ vacant cells. The player will fix the target cell $t_{max}$ and will try to make the same valued tile with respect to it's previous tile. When the value of $t_{max}$ is equal to it's previous tile then the player will merge them. For the favourable situation the player can reach $2^{(\prod_{i=1}^{d}n_i) + 1+ 1}$ easily. By continuing this process $n_2n_3 \ldots n_d$ times the player will reach $2^{(\prod_{i=1}^{d}n_i) + 1+ (\prod_{j=2}^{d}n_j) }$. Then by Induction hypothesis the player can reach the maximum value $2^{(\prod_{i=1}^{d}n_i) + 1}$.
\qed

\begin{corollary}
For the special case $n_1 = n_2 = \ldots =n_d =  n$, that is for the board $\mathcal{G}_{n\times \cdots \times n}$ the maximum value the player can reach is $2^{(n^d+1)}$. \end{corollary}

\pf By Theorem~\ref{thmd}, for  $n_1 = n_2 = \ldots =n_d =  n$ the maximum value the player can reach is $2^{(n^d+1)}$ for \[ \underbrace{n \times n \times \ldots \times n}_{\text{$d$ times}} \] $d$-dimensional board. \qed

\subsection{Algorithm for the $d$ dimensional board $\mathcal{G}_{n_1 \times n_2 \times \ldots \times n_d}$}

We have discussed the algorithm for the two-dimensional board in the previous section. It is very easy to see the method which the computer usees to create a tile on the $d$ dimensional board. So, we are not writing it further. In this section, we have given the algorithm for the player for the board $\mathcal{G}_{n_1 \times n_2 \times \ldots \times n_d}$. As we have written the winning strategy for the player in $d$ dimensions, it will follow that and almost the same as the two-dimensional board algorithm.

\begin{algorithm}[htbp]
\SetAlgoLined
\KwResult{The Player will make the favourable situation to reach the maximum value}
 Choose the direction $\vec{a}$ (see Definition~\ref{def1}) to fix the first target cell in the board $\mathcal{G}_{n_1 \times n_2 \times \ldots \times n_{d-1}}$. \\
 
 \While{The value of $a$ varies over $1, 2, \ldots, d$}{
  calculate the target cells according to the direction \;
  \eIf{there exist tile $t$ with $Val(t)=4$ at $\mathcal{R}_{t_{max}}$}{
  bring tile $t$ at $({i^s}_{{(m,1)}}, {i^s}_{{(m,2)}}, \ldots, {i^s}_{{(m,d)}})$ by one move. \;
   }{
   go to the next step\;
  }
    
    \eIf{there exist tile $t$ with $Val(t)=4$ at $\mathcal{R}_{t_{max}}$ and $t_1, t_2$ with $Val(t_1)=Val(t_2)=2$}{
  Merge $t_1$ and $t_2$ to get $t_3$ with $Val(t_3)=4$. \;
   }{
   Bring tile $t$ with $Val(t) = 2$ at $({i^s}_{{(m,1)}}, {i^s}_{{(m,2)}}, \ldots, {i^s}_{{(m,d)}})$ by one move \;
  }
  
  \eIf{there exist no tile $t$ at $\mathcal{R}_{t_{max}}$}{
  Choose row wise move and bring $t_1$ with $Val(t_1)$ at $\mathcal{R}_{t_{max}}$ from ${\mathcal{R}_{t_{max}}}^{c}$. \;
   }{
   go to the next step \;
  }
  
    \eIf{there exist no tile $t$ with $Val(t) = 4$ in the board $\mathcal{G}_{n_1\times n_2}$}{
  Merge $t_1$ and $t_2$ with $Val(t_1)=Val(t_2)=2$ to get $t_3$ with $Val(t_3)=4$ \;
   }{
   Return to step 4 \;
  }

    }
 
\caption{\textbf{Algorithm for winning strategy of the Player for the board $\mathcal{G}_{n_1 \times n_2 \times \ldots \times n_d}$}}
\label{algomax3}
\end{algorithm}

\newpage

\section{Conclusion}
In this paper, we have analyzed the popular mobile app game `2048' for the first time and generalized it in $n$-dimensional space. Note that Huon Wilson~\cite{cite17}, Cesar Kawakami~\cite{cite18} have developed this result for 4D and 5D respectively. We have also presented a strategy for the computer and a winning strategy for the player under certain conditions.

\end{document}